\documentclass[3p,11pt]{elsarticle}
\usepackage{lineno,hyperref}
\modulolinenumbers[1]
\usepackage{geometry}
\usepackage{graphicx}
\usepackage[usenames,dvipsnames]{color}
\usepackage{color}
\usepackage[english]{babel}
\usepackage{epstopdf}
\usepackage{mathtools}
\usepackage{latexsym}
\usepackage{hyperref}
\usepackage{url}
\usepackage{amstext}
\usepackage{amssymb}
\usepackage{amsmath}
\usepackage{pifont}
\usepackage{siunitx} 
\usepackage{url}
\hypersetup{
    colorlinks=true,
    linkcolor=blue,
    filecolor=magenta,
    urlcolor=blue,
}
\usepackage{times}
\usepackage{hyperref}
\usepackage{url}
\usepackage{amstext}
\usepackage{amssymb}
\usepackage{wrapfig}
\usepackage{float}
\usepackage{tikz}
\usepackage{pgfplots}
\usepackage{caption}
\usepackage{subcaption}
\usepackage[section]{placeins}

\begin{document}

\begin{frontmatter}

    \title{Data-driven Stochastic Model for Quantifying the Interplay Between Amyloid-beta and Calcium Levels in Alzheimer's Disease}

\author[inst1]{Hina Shaheen\corref{cor1}}
\ead{shah8322@mylaurier.ca}
\author[inst1]{Roderick Melnik}
\ead{rmelnik@wlu.ca}
\author[inst2]{Sundeep Singh}
\ead{sunsingh@upei.ca}
\author[inst4]{ The Alzheimer's Disease Neuroimaging Initiative}\ead{ADNI}

\cortext[cor1]{Hina Shaheen}
\address[inst1]{MS2Discovery Interdisciplinary Research Institute, Wilfrid Laurier University, Waterloo, ON N2L 3C5, Canada}
\address[inst2]{Faculty of Sustainable Design Engineering, University of Prince Edward Island, Charlottetown, PE C1A 4P3, Canada}
\address[inst4]{Data used in preparation of this article were generated by the Alzheimer’s Disease Metabolomics Consortium (ADMC). As such, the investigators within the ADMC provided data, but did not participate in the analysis or writing of this report. A complete listing of ADMC investigators can be found at: \url{https://sites.duke.edu/adnimetab/team/}}

\begin{abstract}
The abnormal aggregation of extracellular amyloid-$\beta$ ($A\beta$) in senile plaques resulting in calcium ($Ca^{+2}$) dyshomeostasis is one of the primary symptoms of Alzheimer's disease (AD). Significant research efforts have been devoted in the past to better understand the underlying molecular mechanisms driving $A\beta$ deposition and $Ca^{+2}$ dysregulation. Importantly, synaptic impairments, neuronal loss, and cognitive failure in AD patients are all related to the buildup of intraneuronal $A\beta$ accumulation. Moreover, increasing evidence show a feed-forward loop between $A\beta$ and $Ca^{+2}$ levels, i.e. $A\beta$ disrupts neuronal $Ca^{+2}$ levels, which in turn affects the formation of $A\beta$. To better understand this interaction, we report a novel stochastic model where we analyze the positive feedback loop between $A\beta$ and $Ca^{+2}$ using ADNI data. A good therapeutic treatment plan for AD requires precise predictions. Stochastic models offer an appropriate framework for modelling AD since AD studies are observational in nature and involve regular patient visits. The etiology of AD may be described as a multi-state disease process using the approximate Bayesian computation method. So, utilizing ADNI data from $2$-year visits for AD patients, we employ this method to investigate the interplay between $A\beta$ and $Ca^{+2}$ levels at various disease development phases. Incorporating the ADNI data in our physics-based Bayesian model, we discovered that a sufficiently large disruption in either $A\beta$ metabolism or intracellular $Ca^{+2}$ homeostasis causes the relative growth rate in both $Ca^{+2}$ and $A\beta$, which corresponds to the development of AD. The imbalance of $Ca^{+2}$ ions causes $A\beta$ disorders by directly or indirectly affecting a variety of cellular and subcellular processes, and the altered homeostasis may worsen the abnormalities of $Ca^{+2}$ ion transportation and deposition. This suggests that altering the $Ca^{+2}$ balance or the balance between $A\beta$ and $Ca^{+2}$ by chelating them may be able to reduce disorders associated with AD and open up new research possibilities for AD therapy.

\end{abstract}


\end{frontmatter}

\section{Introduction}
Alzheimer's disease (AD) is the most prevalent kind of adult dementia. AD is a medical disorder that gradually kills neurons and produces severe cognitive impairment \cite{koper2020necrosome,gomez2022lesions}. Many medication therapies are shown to slow the progression of AD, but there is no permanent cure \cite{breijyeh2020comprehensive,grutzendler2001cholinesterase,vaz2020alzheimer,nguyen2021amyloid,toro2016brain}. The clinical and pathological hallmarks of AD include progressive neuronal loss, synaptic degradation, and the formation of amyloid plaques and neurofibrillary tangles in particular regions of the brain \cite{shaheen2021neuron,pal2022influence}. According to some findings, AD may be a systemic disease since it affects not only neurons but also peripheral cells such as fibroblasts, lymphocytes, and platelets in AD patients \cite{trushina2019alzheimer,ullah2021abnormal,bell2020peripheral}.

Although the exact cause of AD is unknown, a few major theories, including the cholinergic, amyloid cascade, and tau hypothesis, have been presented to explain the progression of AD. The amyloid cascade theory appears to be the most likely, as there are numerous plaques composed of amyloid ($A\beta$) peptide in the AD patient's brain \cite{castellani2019amyloid,ashleigh2023role}. According to the amyloid cascade theory, $A\beta$ oligomers or amyloid fibrils are formed by the aggregation of $A\beta$ oligomers or amyloid fibrils, which are key components of $A\beta$ peptide and impair the function of neuronal cells \cite{shaheen2021neuron}. Many mathematical models have been presented to explain the development of $A\beta$ monomer synthesis or aggregation (see \cite{pal2022influence,pal2022coupled,fornari2020spatially,fornari2019prion} and the references therein). Another hypothesis that has attracted a lot of attention, proposes that disturbance of calcium ($Ca^{+2}$) homeostasis is crucial to AD pathogenesis \cite{cascella2021calcium,alzheimer2017calcium,yu2009calcium}. The disruption of $Ca^{+2}$ homeostasis has been extensively explored in order to understand the processes of $A\beta$-induced neurotoxicity. Intracellular $Ca^{+2}$ operates as a second messenger, regulating neuronal activities such as brain development and differentiation, action potential, and synaptic plasticity \cite{shaheen2021neuron,genovese2020sorcin,jadiya2019impaired}. The $Ca^{+2}$ hypothesis of AD proposes that activation of the amyloidogenic pathway affects neuronal $Ca^{+2}$ homeostasis as well as the processes involved in learning and memory. $A\beta$ may alter $Ca^{+2}$ signalling by numerous methods, including boosting $Ca^{+2}$ inflow from the extracellular area and stimulating $Ca^{+2}$ release from intracellular repositories within the brain \cite{sehar2022amyloid,calvo2021mitochondria}. Moreover, growing evidence suggests that there is a positive loop between $Ca^{+2}$ and $A\beta$ levels \cite{zhang2020mathematical,muller2023dynamics,cristovao2019s100}. We know by now, for example, that the persistent high concentration of $Ca^{+2}$ is favorable to the formation of $A\beta$ in rat cortical neurons by imitating $A\gamma$ secretase activity, which is crucial for the breakdown of amyloid precursor protein (APP) \cite{zhang2020mathematical,tong2018calcium}.

Some of the causes of substantial trial failure include an inadequate understanding of AD etiology and development, as well as an inappropriate trial design. The creation of clinical trial simulations and mathematical modelling of AD progression are key tools for exploring the reasons why clinical trials fail and refining the clinical trial methodology. The poorly understood nature of AD etiology and development limits the capacity to build solid mechanistic models for reliable disease progression prediction. There are also mathematical models based on inverse problems that have been established to reflect modifications to cognition over time, as measured by errors on various cognitive tests used to assess patients' intellectual capabilities, such as the Modified Mini-Mental State Examination (MMSE) and the AD Assessment Scale \cite{ashford2001modeling,rogers2012combining,samtani2012improved,ito2011disease,mar2015fitting,hadjichrysanthou2018development,veitch2022using,hao2022optimal}. The majority of models developed to date in the aforesaid context are stochastic in nature, as evidenced by \cite{zhang2020mathematical,lorenzi2019probabilistic,green2011model}. The FDA has authorized following  medicines for the clinical treatment of AD, i.e., tacrine, rivastigmine, galantamine, and donepezil, which are acetylcholinesterase inhibitors (AChEIs), which increase the concentration and duration of the neurotransmitter acetylcholine's activity (Ach) and another therapeutically utilized medicine is memantine, which is an N-Methyl-d-aspartate (NMDA) receptor antagonist \cite{dong2019drug}. Additionally, lecanemab and aducanumab are newly approved medicines. In the brains of AD patients, NMDA receptors are overstimulated due to glutamate excess release by neurons, resulting in increased intracellular $Ca^{+2}$ and the death of neuronal cells. Therefore, reducing the concentrations of $Ca^{+2}$ fluxes during the disease state could stop the progression of AD. 

Importantly, a persistent increase in baseline $Ca^{+2}$ may also play a role in disease progression by increasing the synthesis and toxicity of $A\beta$'s in cells harboring AD-related mutations. These mutually beneficial regulations, in which $A\beta$ promotes a $Ca^{+2}$ increase, which in turn raises the level of $A\beta$, create a positive feedback loop that is expected to create a vicious circle leading to disease development \cite{shaheen2021neuron,zhang2020mathematical,de2013progression}. Inspired by this fact and using clinical data such as study data from the ADNI database (\url{adni.loni.usc.edu}), we developed and tested a simple novel stochastic model to predict the interplay between the $A\beta$ and $Ca^{+2}$ concentrations on AD progression in a clinical trial. Also, we have selected the AD patients to be monitored at frequent visits, i.e., between $0-2$ year visits. We analyzed the data for $A\beta$ concentration and fitted it to the developed stochastic model using the approximate Bayesian computation (ABC) technique. ABC is a data-driven strategy that utilizes a number of low-cost numerical simulations.
ABC evaluates unknown physical or model parameters and associated uncertainties using reference data from real-world experiments or higher-fidelity numerical simulations \cite{beaumont2019approximate}. We found that during the disease state in the patient's brain, there is a tremendous increase in $A\beta$ oligomers, which enhance the influxes of intracellular $Ca^{+2}$. In return, $Ca^{+2}$ encourages the production of these hazardous $A\beta$ oligomers, and this fact reinforces the positive feedback between $Ca^{+2}$ and $A\beta$. We show that the simulations of our model with the ADNI data correlate with the finding that a variety of dysregulations in $Ca^{+2}$ and $A\beta$ may lead to disease, as well as random fluctuations of $A\beta$ in vulnerable patients that can lead to a transition from the ``healthy'' to the ``pathological'' state \cite{shaheen2021neuron,pal2022coupled,zhang2020mathematical,de2013progression,hao2016mathematical,andrade2013cell,demuro2011single,kuchibhotla2008abeta,berridge2016inositol,o2023calmodulin}. This vulnerability may explain the high prevalence of sporadic AD found in the elderly population. 

The rest of the paper is organized as follows: In Section \ref{meth}, we describe the modelling approach based on the (i) deterministic and stochastic models of the interplay between $A\beta$ and $Ca^{+2}$, and (ii) the ABC technique. In Section \ref{exp}, we set up the experimental data and model the participant dynamics. We present our results and computational simulations based on the developed stochastic models in Section \ref{res}. The computational results have been obtained with an in-house developed MATLAB code, and all data analysis has been carried out in Python. Finally, we discuss our results, conclude our findings, and outline future directions in Sections \ref{dis} and \ref{con}.

\section{Methodology}\label{meth}
This section highlights the deterministic and stochastic modelling approaches for AD incorporating the interplay between $A\beta$ and $Ca^{+2}$ within the Bayesian setting. This is achieved by defining the stochastic model of $A\beta$ and $Ca^{+2}$ by adding the stochastic noises. Additionally, we simulate the trajectories of the stochastic model of $A\beta$ and $Ca^{+2}$ using the ADNI data by incorporating the ABC technique presented in section \ref{abc}. The aim is to fit the ADNI data of $A\beta$ concentrations into the stochastic model of $A\beta$ and see the impact on  $Ca^{+2}$ concentrations. 

\subsection{The modelling approach}\label{sectionm}
In this section, we will present the mathematical model for AD to account for the coexistence between  $Ca^{+2}$ and A$\beta$.
\begin{figure}[htbp]
\centering
\includegraphics[scale=.5]{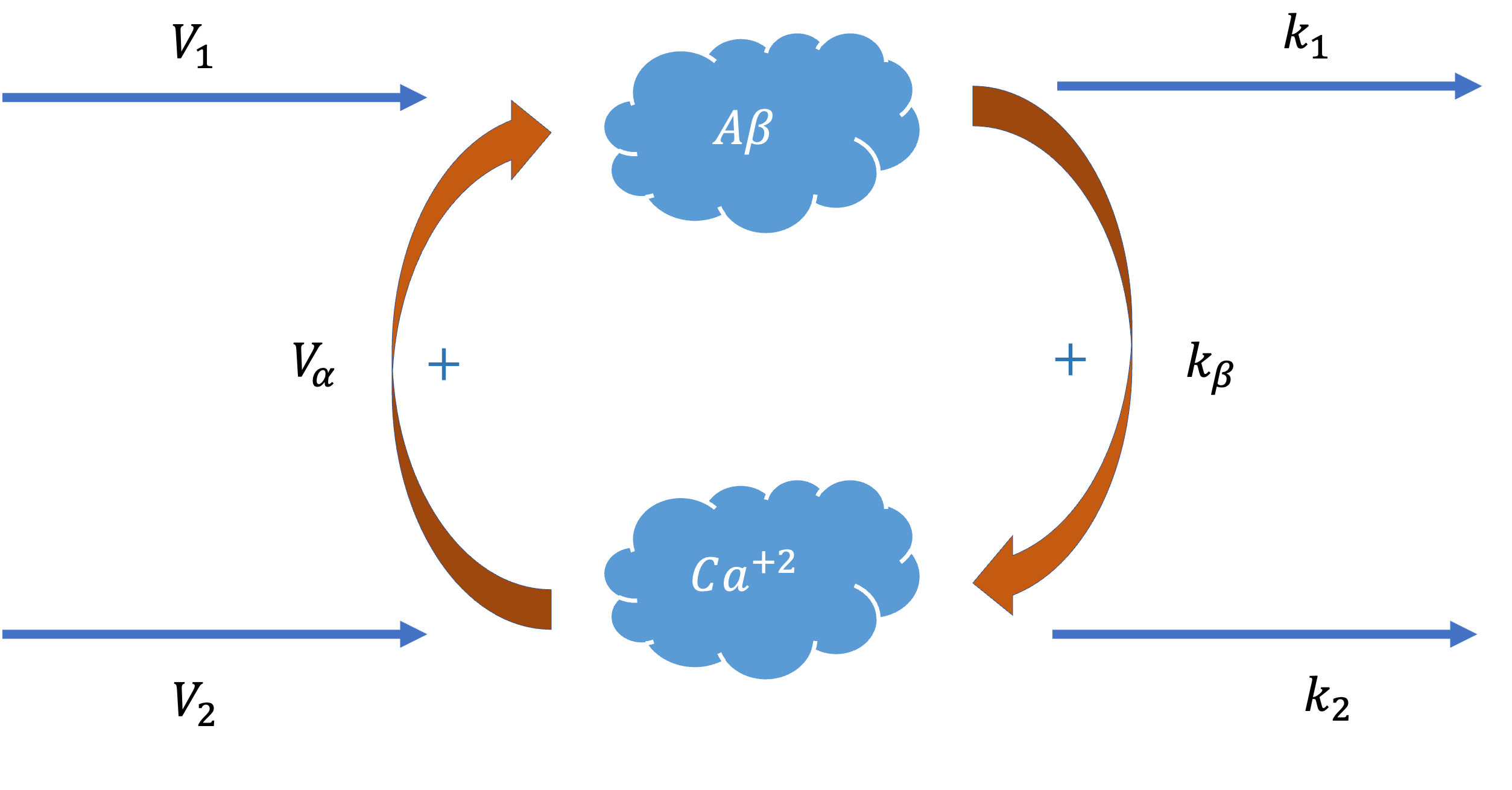}
\caption{(Color online) Graphic illustration of the mathematical model provided by Eqs. (\ref{eq1})-(\ref{eq2}) that describes the interaction between $Ca^{+2}$ and $A\beta$ during the progression of AD. The positive feedback exerted by $Ca^{+2}$ on the creation of $A\beta$, as well as the fact that $A\beta$ tends to increase intracellular $Ca^{+2}$, establish a positive loop (motivated by \cite{de2013progression}).}
\label{fig:1}       
\end{figure}
AD is associated with A$\beta$ produced by the cleavage of the amyloid precursor protein (APP), which is partly embedded in the plasma membrane. APP is cleaved by either an $\alpha$- or a $\beta$-secretase. In the amyloidogenic pathway, cleavage of APP by the $\beta$-secretase generates sAPP$\beta$ and CTF$\beta$. The latter is in turn cleaved by a $\gamma$-secretase to form $A\beta$. A rise in cytosolic $Ca^{+2}$ enhances the production and release of A$\beta$ which leads to stimulation of $\gamma$-secretase activity in cortical neurons \cite{de2013progression}. In resting neurons, the free cytosolic $Ca^{+2}$ level is maintained around $50-100\si{nM}$, while it is increased up to $1\si{\mu M}$ upon electrical or receptor-mediated stimulation. As described in \cite{shaheen2021neuron}, $Ca^{+2}$ influx is enhanced by VGCCs or ligand-gated ion channels such as glutamate and acetylcholine receptors. However, the main intracellular $Ca^{+2}$ store is the endoplasmic reticulum(ER), from where $Ca^{+2}$ can be released through the inositol $1,4,5$-trisphosphate receptor ($IP_{3}R$) or through the ryanodine receptor. Decrease of cytosolic $Ca^{+2}$ occurs through $Ca^{+2}$-ATPases, the $Na^{+}/Ca^{2+}$ exchanger, or the mitochondrial uniporter. $A\beta$'s perturb the balance between $Ca^{+2}$ entry into and extrusion out of the cytoplasm. In healthy neurons, this process equilibrates, leading to a basal $Ca^{+2}$ level in the range of $50–100 \si{nM}$. Using transgenic mouse models for AD together with $Ca^{+2}$ imaging, Kuchibhotla et al. \cite{kuchibhotla2008abeta} have shown that this resting concentration is higher in neurites located close to amyloid deposits, while another study reports that the basal level of $Ca^{+2}$ in the cortical neurons of such animals is around $250\si{nM}$, i.e. twice that found in controls \cite{lopez2008increased}. $Ca^{+2}$ channels are also deregulated in brain cells and the formation by $A\beta$ oligomers of pores in the plasma membrane enhances the influx of extracellular $Ca^{+2}$ \cite{tong2018calcium}. This feedback is further reinforced by the fact that $Ca^{+2}$ promotes the formation of these toxic oligomers.

We will propose the simple model schematized in Fig. \ref{fig:1} based on the experimental observations using ADNI data explained in Section \ref{exp}. The main variables of the model are the intracellular $Ca^{+2}$ concentrations and the concentrations of $A\beta$ (without distinction between intracellular and extracellular compartments, nor between amyloid compounds of different lengths and in different oligomerization states). These concentrations are denoted by $A\beta$ and $Ca$ in the equations, respectively. 
The evolution of the two variables of the model equation is given as follows:
\begin{equation}\label{eq1}
    \frac{d(A\beta)}{dt}=V_1+V_{\alpha}\frac{(Ca)^n}{K_{\alpha}^n+(Ca)^n}-k_1(A\beta),
\end{equation}
\begin{equation}\label{eq2}
    \frac{d(Ca)}{dt}=V_2+k_{\beta}(A\beta)^m-k_2(Ca),
\end{equation}
where the $Ca$ concentration represents the basal level of cytoplasmic $Ca^{+2}$, whose value does not significantly depend on the short-lived $Ca^{+2}$ peaks arising from the electrical activity of the neurons. $A\beta$ is assumed to be synthesized at a constant rate $V_1$ and eliminated with first-order kinetics, characterized by a rate constant $k_1$. Activation of $A\beta$ synthesis by $Ca^{+2}$ is represented by a Hill term with a maximal rate of $V_{\alpha}$, a half-saturation constant of $K_{\alpha}$ and a Hill coefficient $n$. Similarly, $Ca^{+2}$ enters the cytoplasm at
a constant rate $V_2$ and is eliminated with first-order kinetics, characterized by a rate constant $k_2$. Moreover, $A\beta$  oligomers induce $Ca^{+2}$ entry into the cell, putatively by provoking an increase in plasma membrane permeability. This process is characterized by a cooperativity coefficient $m$, and a rate constant $k_\beta$. This latter term is taken as non-saturable to model the formation of pores by oligomers of $A\beta$. The model Eqs. (\ref{eq1})-(\ref{eq2}) are adopted from \cite{de2013progression}, It is considered that in a healthy neuron, the concentrations of $A\beta$ ($\sim5$ \si{nM}) are lower than those of $Ca^{+2}$ ($\sim50-100$ \si{nM}) \cite{sheng2012synapses}. In the present section, we formalize this positive loop in a mathematical model and show that it exhibits bistability. Therefore, a stable steady state characterized by low levels of $Ca^{+2}$ and amyloids, which correspond to a healthy situation, coexists with another ‘pathological state’ where the levels of both compounds are high. The onset of the disease corresponds to the switch from the lower steady-state to the higher one induced by a large enough perturbation in either the metabolism of amyloids or the homeostasis of intracellular $Ca^{+2}$. 

It is well known that the majority of models created to date for AD using clinical data in the above context are stochastic in nature \cite{hadjichrysanthou2018development,lorenzi2019probabilistic,green2011model,bilgel2019predicting,burnham2020impact}. Such models have the important advantage of allowing for variation in model parameters and disease biomarkers when predicting disease progression. Therefore, stochastic noises can be incorporated into the present model of AD i.e., Eqs (\ref{eq1})-(\ref{eq2}), which focuses on the evolution of $A\beta$ and $Ca^{+2}$. The dynamics of $A\beta$ and $Ca^{+2}$ are perturbed by intrinsic or extrinsic noises. The intrinsic noises arise from the random fluctuations of biochemical reaction events such as the stimulation of calcium on $\gamma$-secretase activity, the nucleated aggregation process, and the changes of cell membrane integrity induced by $A\beta$ \cite{eugene2016insights}, whereas the extrinsic noises originate from the stochastic variations of the microenvironment for $A\beta$ and $Ca^{+2}$, which include $pH$, the concentrations of $Na^{+}$, reactive oxygen species, neurons, and peripheral macrophages \cite{de2013progression}. Additionally, AD is a neurological disorder that progresses over a long period of time, from a normal state to severe dementia. In contrast, the concentration of $Ca^{+2}$ changes quickly, which is at the timescale of seconds or minutes \cite{zhang2020mathematical}.

On the basis of the above biological backgrounds, we will incorporate the stochastic noises and the explicit time scales into the model Eqs. (\ref{eq1})-(\ref{eq2}) as follows:

\begin{equation}\label{eq3}
    d(A\beta)=\big(V_1+V_{\alpha}\frac{(Ca)^n}{K_{\alpha}^n+(Ca)^n}-k_1(A\beta)\big)dt+ \sigma_1\sqrt{\epsilon}(A\beta) dB_1(t),
\end{equation}
\begin{equation}\label{eq4}
    d(Ca)=(V_2+k_{\beta}(A\beta)^m-k_2(Ca))dt+\sigma_2(Ca)dB_2(t),
\end{equation}
where $0<\epsilon<<1$ is used to indicate that the change of $A\beta$ concentration is much slower than that of $Ca^{+2}$ concentration and $B_i(t), i = 1,2,$ represent the standard Wiener process defined on a complete probability space $(\Omega,\mathcal{F},\mathcal{P})$ and $\sigma_i^2 > 0$ for $i = 1,2,$ denote the intensities of white noise, other relevant parameters are adopted from \cite{zhang2020mathematical}. The aim of adding a stochastic term to the model is to show that the stochastic noises can induce a jump transition from a state with a lower concentration of $A\beta$ to a state with a higher concentration of $A\beta$ using ADNI data. Such jump transactions represent a key phenomenon for AD. Secondly, we analyze the impacts of stochastic noises on the progression of $A\beta$ and $Ca^{+2}$ since AD models are stochastic. The novelty of the present research is in the development of a stochastic AD model of $A\beta$ and $Ca^{+2}$ using ADNI data for AD patients at $2$-year visits (details are given in Section \ref{exp}). It is expected that numerical simulations of the model will reproduce a variety of experimental observations about the disease using the ADNI data, which could be useful when developing therapeutic protocols to slow down the progression of AD. Since the ADNI data contains only the concentrations of $A\beta$, thus the aim is to fit the $A\beta$ concentration data and incorporate its effect on $Ca^{+2}$ concentrations within the developed stochastic models. This can be done using the ABC technique with details given in Section \ref{abc}. 

\begin{table}
\centering
\caption{List of the parameter values used for deterministic and stochastic models adopted from \cite{de2013progression}.}
\label{tab:1}       
\fontsize{10pt}{10pt}
%
%
\begin{tabular}{ |p{3cm}|p{5cm}| }
\hline
Parameters & Values (units) \\
\hline
$V_1$ &  $0.0065(\si{nM}/year)$\\\hline
$V_{\alpha}$ &  $0.05(1/year)$\\\hline
$K_{\alpha}$&   $120\si{nM}$\\\hline
$n$ &  $2$\\\hline
$k_1$ &  $0.01(1/year)$\\\hline
$V_2$ & $ 5(\si{nM}/year)$\\\hline
$k_{\beta}$ &  $2*10^{-1}(\si{nM}/year)$\\\hline
$m$ &  $4$\\\hline
$k_2$& $1*10^{-1}(1/year)$\\
\hline
\end{tabular}
\end{table}
\subsection{Approximate Bayesian Computation (ABC) technique} \label{abc}
ABC is a data-driven approach that employs several low-cost numerical simulations. Using reference data from real-world experiments such as ADNI data or higher-fidelity numerical simulations, ABC also estimates unknown physical or model parameters, as well as their uncertainties \cite{christopher2018parameter}. 

Bayesian inference allows for the estimation of instability by evaluating the likelihood of the model parameters supplied by the experimental data \cite{beaumont2019approximate,calvetti2018inverse,iftekharuddin2009fractal}. Since our template contains only the $A\beta$ concentrations data, we treat the model Eqs. (\ref{eq3})-(\ref{eq4}) as inverse problems and get data for ``$A\beta$" from the ADNI database for AD patients per $2$-year visit. To solve the differential equations (i.e., Eqs. \ref{eq3}-\ref{eq4}) using Bayesian inference, we need to first specify the prior distributions for the unknown parameters (i.e., $V_1$, $V_{\alpha}$, $K_{\alpha}$, $n$, $k_1$, $V_2$, $k_{\beta}$, and $m$), and then use Bayesian methods to update these priors based on the observed data. Let's assume that the priors for all the parameters are independent and normally distributed with a mean of $0$ and a variance of $10$. This is a fairly non-informative prior that allows for a wide range of possible values for the parameters. Next, we need to define the likelihood functions for the two differential equations. The likelihood function for the first equation (i.e., Eq. \ref{eq3}) is given by:
\begin{equation}
    p(\mathbf{A\beta}|\mathbf{\theta}_1,t) = \prod_{i=1}^{n}\frac{1}{\sqrt{2\pi\sigma^2}}\exp\left(-\frac{(A\beta_i-\widehat{{A\beta_i}})^2}{2\sigma^2}\right),
\end{equation}
where $\mathbf{A\beta}=(A\beta_1,A\beta_2,\ldots,A\beta_{n})$ is the vector of observed values for $A\beta$ at times $t=(t_1,t_2,\ldots,t_{n})$, $\widehat{{A\beta}_i}$ is the predicted value of $\mathbf{A\beta}$ at time $t_i$ based on the current parameter values $\mathbf{\theta}_1$, and $\sigma^2$ is the measurement error variance. The predicted values $\widehat{{A\beta}_i}$ can be obtained by numerically solving the differential equation using the current parameter values as given in Table \ref{tab:1}. We can use any numerical solver, such as the Runge-Kutta method, to do this. The likelihood function for the second equation (i.e. Eq. \ref{eq4}) is given by:
\begin{equation}
    p(\mathbf{Ca}|\mathbf{\theta}_2,t) = \prod_{i=1}^{n}\frac{1}{\sqrt{2\pi\sigma^2}}\exp\left(-\frac{(Ca_i-\widehat{Ca}_i)^2}{2\sigma^2}\right),
\end{equation}
where $\mathbf{Ca}=(Ca_1,Ca_2,\ldots,Ca_{n})$ is the vector of observed values for $Ca$ at times $t=(t_1,t_2,\ldots,t_{n})$, $\widehat{Ca}_i$ is the predicted value of $\mathbf{Ca}$ at time $t_i$ based on the current parameter values $\mathbf{\theta}_2$, and $\sigma^2$ is the measurement error variance. The predicted values $\widehat{Ca}_i$ can also be obtained by numerically solving the differential equation (i.e., Eq. \ref{eq4}) using the current parameter values. To update the priors based on the observed data, we use Bayes' theorem \cite{berrar2018bayes}:
\begin{equation}
    p(\mathbf{A\beta},\mathbf{Ca}|\mathbf{\theta},t) \propto p(\mathbf{A\beta}|\mathbf{\theta}_1,t)p(\mathbf{Ca}|\mathbf{\theta}_2,t)p(\mathbf{\theta})
\end{equation}
where $\mathbf{\theta}=(V_1,V_{\alpha},K_{\alpha},n,k_1,V_2,k_{\beta},m)$ is the vector of unknown parameters. We use a Markov Chain Monte Carlo (MCMC) algorithm, a Hamiltonian Monte Carlo approach implemented in the Python package PyMC3 to sample from the posterior distribution $p(\mathbf{A\beta},\mathbf{Ca},t|\mathbf{\theta})$ and obtain estimates of the posterior mean \cite{wang2017bayesian,salvatier2016probabilistic}. After inserting the sampled parameters into our model and comparing the resulting simulated uptake values of $\mathbf{A\beta}$ and $\mathbf{Ca}$ with the observed data, we can rate each sample based on its likelihood and use Bayes' theorem to determine the posterior distributions of most likely parameter values for each patient. To fit the ADNI data into the developed stochastic models, we replace the observed data with ADNI data for $A\beta$ concentrations and the time $t$, as well as the age of the patients given in years. The details on the ADNI experimental setup data are given as follows. 

\section{Experimental setup analysis supported by ADNI data}\label{exp}
\subsection{ADNI data}
The datasets used in the preparation of this article were obtained from the Alzheimer's Disease Neuroimaging Initiative (ADNI) database( \url{adni.loni.usc.edu}). The primary goal of the ADNI has been to test whether serial magnetic resonance imaging (MRI), positron emission tomography (PET), other biological markers, and clinical and neuropsychological assessment can be combined to measure the progression of AD. Specifically, we used the ADNI data prepared for the AD modelling challenge and followed the recommendations prescribed in this work to incorporate the interplay between $A\beta$ and $Ca^{+2}$ using the stochastic approach described in Section \ref{meth}, and fit data for the progression of AD. 
\subsection{Modelling the participant's dynamics}
We use the qualitative template (available as a CSV file by permission on the ADNI website, (\url{adni.loni.usc.edu})) for the progression of the AD project dataset, which includes the three ADNI phases: ADNI 1, ADNI GO, and ADNI 2. This dataset contains measurements from brain MRI, PET, CSF, cognitive tests, demographics, and genetic information \cite{bilgel2019predicting}. From ADNI 1/GO/2, we used the data for $1706$ individuals with $6880$ visits. The ADNI Conversion Committee made clinical diagnoses of MCI (mild cognitive impairment), NL (normal), EMCI (early mild cognitive impairment), LMCI(late mild cognitive impairment), and AD based on the standards outlined in the ADNI protocol. We designed an analysis based on the data for the individuals with clinical follow-up visits to fall within the $24$-month window. Moreover, in the data we use, clinical follow-up visits with improperly arranged dates are discarded for each individual. Also, based on the data statistics, an acceptable time gap of $12$ months is estimated for the present study. In the data set, measurements and clinical diagnoses with missing dates and information per visit are set to missing values in order to use the suggested procedure, and participants with fewer than two distinct visits are excluded. Notably, data sets containing missing values and clinical status are denoted as ``Missing". For the present study, we have used only the AD patients' data. We assume that the participants with AD brains for each visit between $2$ years have different $\mathbf{A\beta}$ concentrations as obtained from the ADNI. Also, we choose the mean of the baseline (bl), $12$ and $24$ months values of $\mathbf{A\beta}$ concentration as the true initial value of $\mathbf{A\beta}$, i.e., $A\beta_0$ to incorporate the Bayesian inference described in Section \ref{abc}. Therefore, using these initial values first, we will fit the ADNI data to the developed stochastic models, i.e., Eqs. (\ref{eq3})-(\ref{eq4}). Then, we will investigate the relationship between $\mathbf{A\beta}$ and $\mathbf{Ca}$ concentrations by fitting the ADNI data into the developed stochastic models. For the present study, we simply considered only AD patients, given our focus on examining the interplay between $\mathbf{A\beta}$ and $\mathbf{Ca}$ concentrations during the disease state. 

\section{Results}\label{res}
\begin{figure}[htbp]
\centering
\includegraphics[scale=.8]{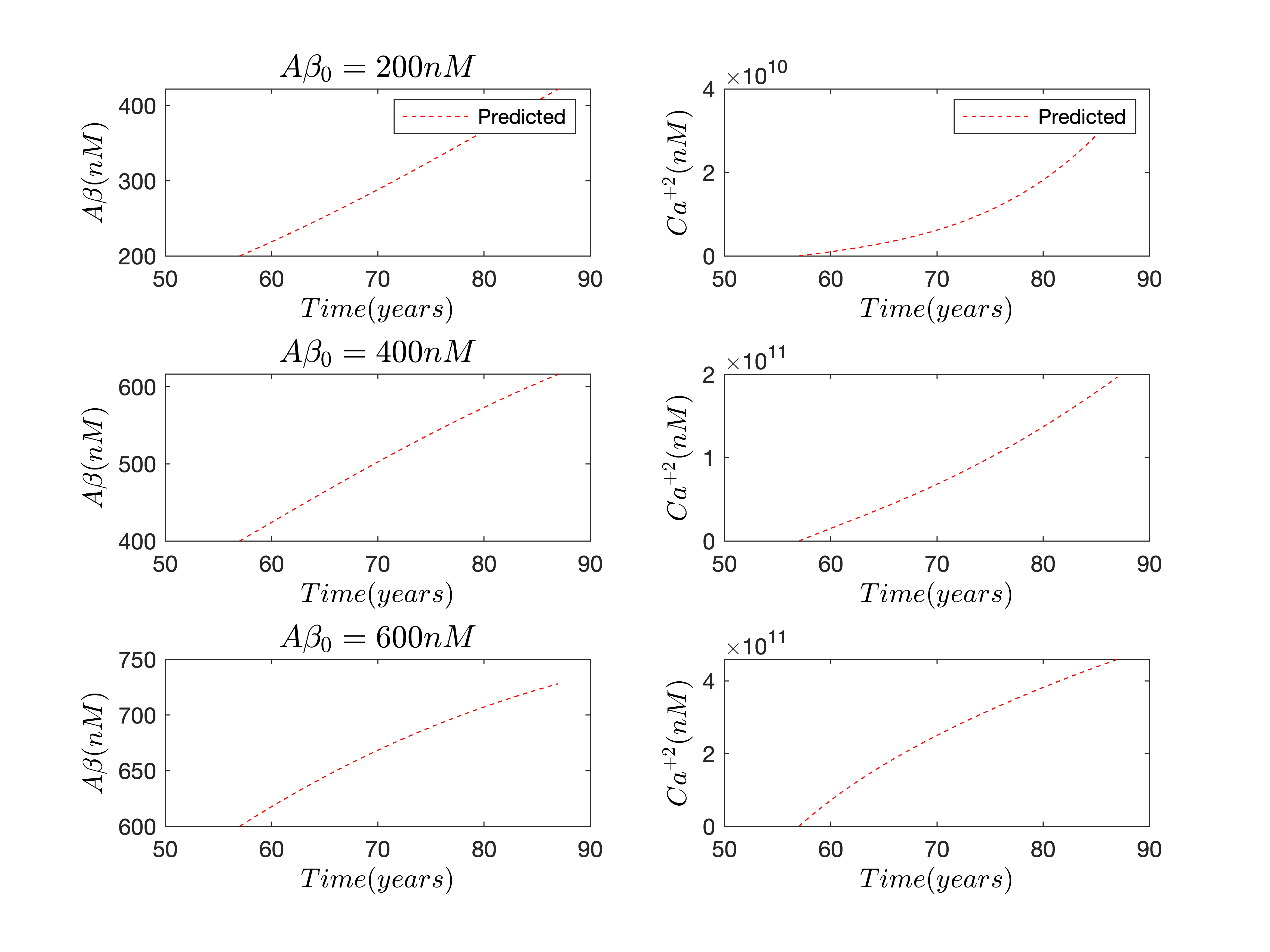}
\caption{(Color online) The trajectories of the stochastic model by fitting the simulated data describing the interplay between $Ca^{+2}$ and $A\beta$ during the onset of AD and defined by
Eqs. (\ref{eq3})-(\ref{eq4}). The initial values of $A\beta$ chosen here are $200,400,600 \si{nM}$ (top to bottom) and for $Ca^{+2}$, set it to $0$. Again, the concentration of $A\beta$ is defined by time scales on a number of years, whereas $Ca^{+2}$ is characterized by seconds to minutes. These values are chosen based on our assumptions to analyze ADNI data, also, $A\beta$ is a promising biomarker that is measured in CSF fluids collected from ADNI participants measured in $\si{nM}$}
\label{fig:4}       
\end{figure}

\begin{figure}[htbp]
\centering
\includegraphics[scale=.8]{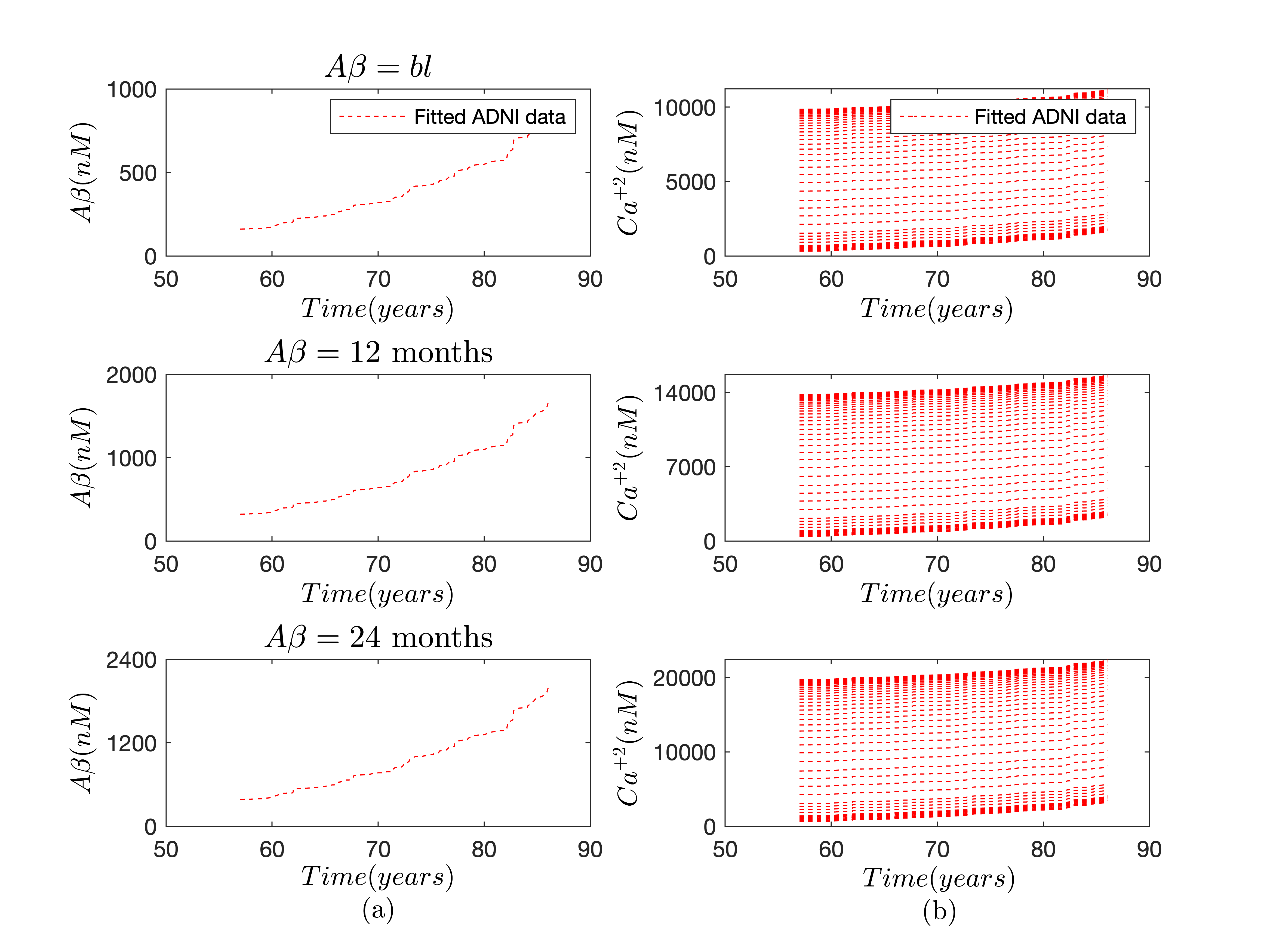}
\caption{(Color online) The trajectories of the stochastic model by fitting the ADNI data describing the interplay between $Ca^{+2}$ and $A\beta$ during the onset of AD and defined by
Eqs. (\ref{eq3})-(\ref{eq4}). The initial values of $A\beta$ chosen here are the means of $A\beta$ concentrations at the bl, $12$ and $24$ month visits (from top to bottom) which were taken as $200\si{nM},400\si{nM},600\si{nM}$. (a) The impact of $A\beta$ (left) on (b) $Ca^{+2}$ (right) is presented. The initial values of $Ca^{+2}$ are set to zero. }
\label{fig:5}       
\end{figure}

\begin{figure}[htbp]
     \centering
     \begin{subfigure}[b]{0.45\textwidth}
         \includegraphics[width=\textwidth]{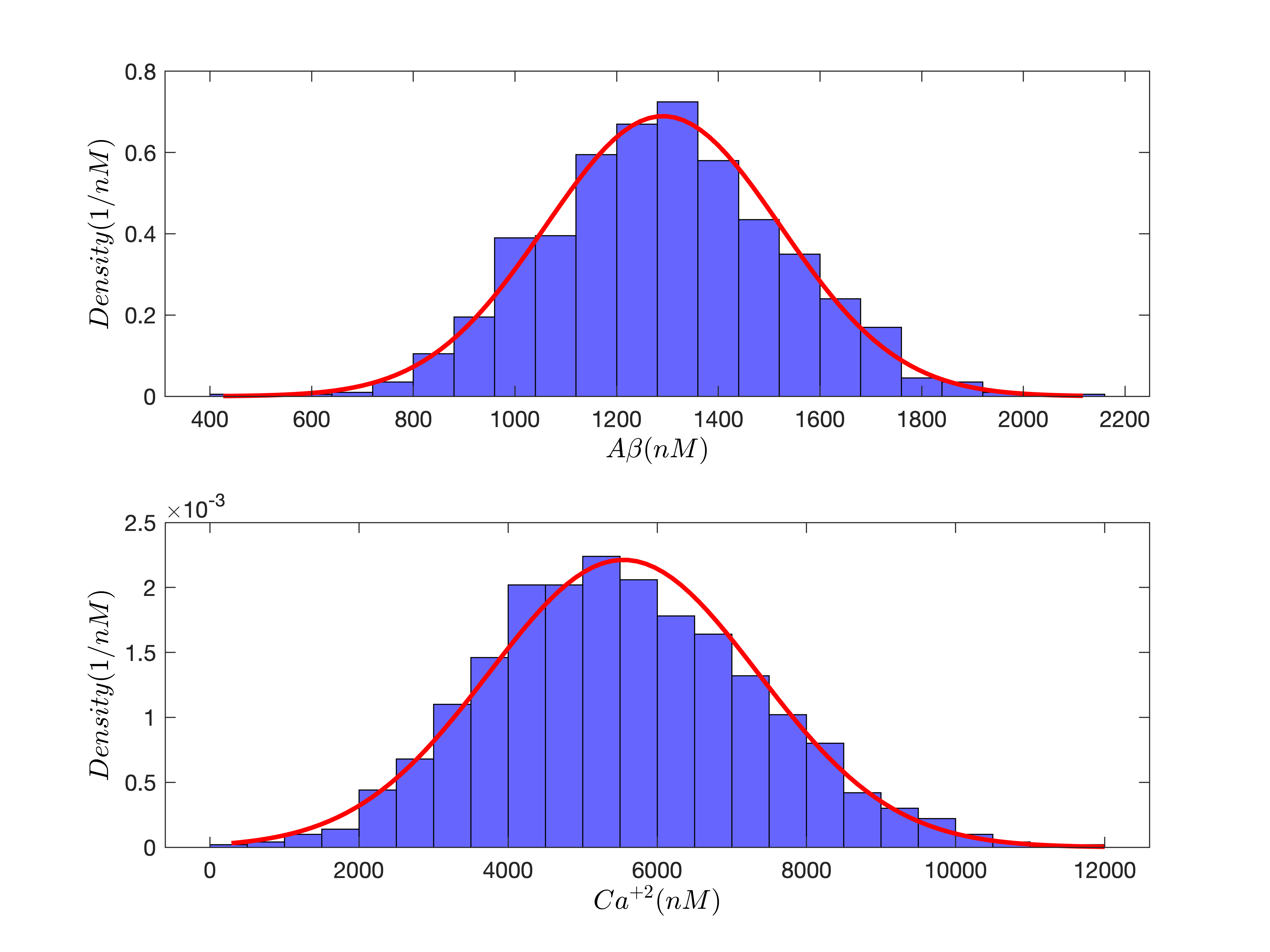}
         \caption{}
   \label{foura}
     \end{subfigure}
     \begin{subfigure}[b]{0.45\textwidth}
         \includegraphics[width=\textwidth]{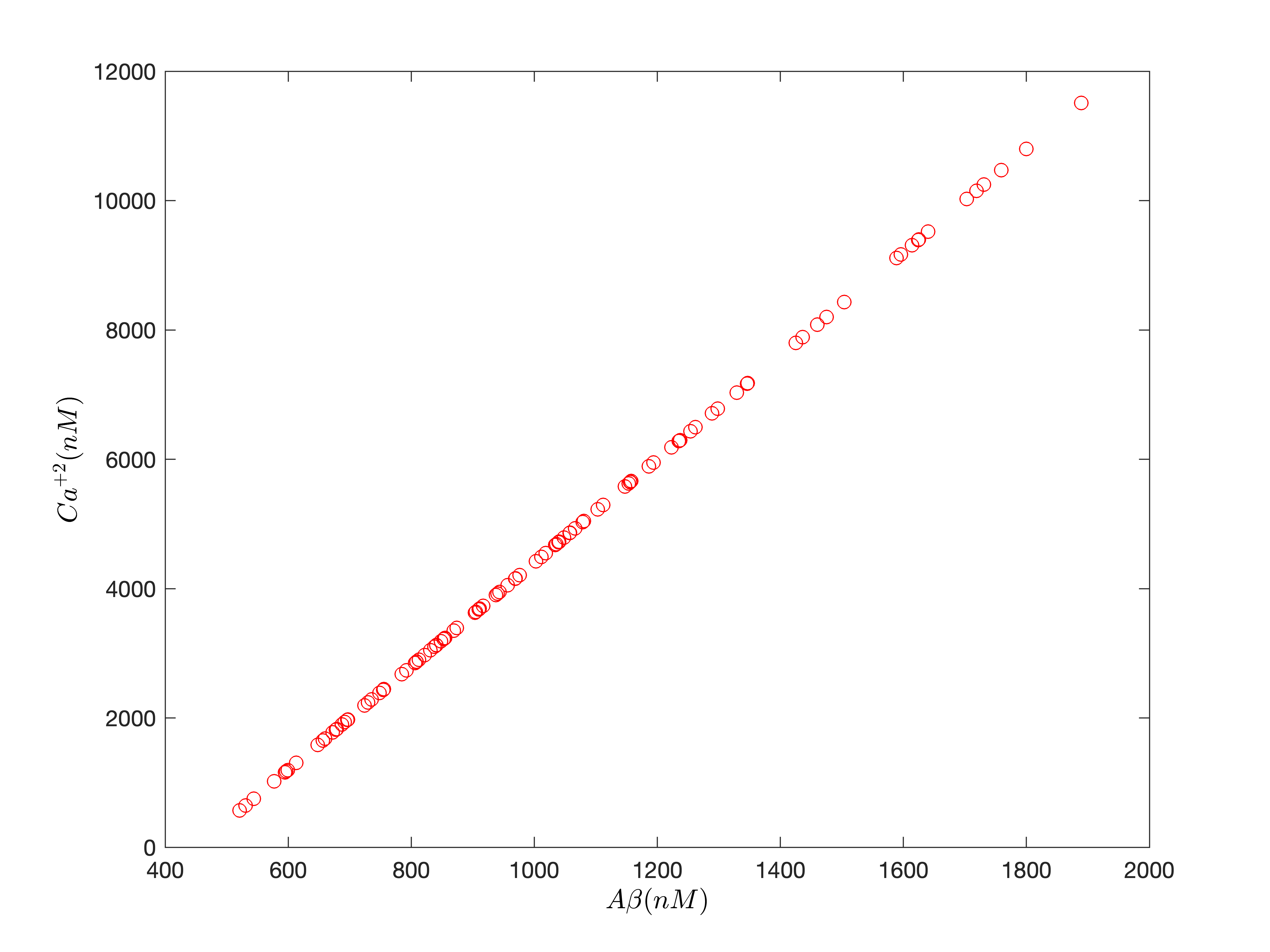}
         \caption{}
       \label{fourb}
     \end{subfigure}
        \caption{(Color online) (a) The blue histograms represent the stationary density of $A\beta$ (top) and $Ca^{+2}$ (bottom) obtained by stochastic simulations by fitting the ADNI data to the Eqs. (\ref{eq3})-(\ref{eq4}), the red lines stand for stationary density in Eqs. (\ref{eq3})-(\ref{eq4}), the y-axis represents the probability density function ($1/nM$). (b) The straight line represents the interplay between $A\beta$ and $Ca^{+2}$ concentrations after fitting the ADNI data into Eqs. (\ref{eq3})-(\ref{eq4}) The initial conditions for for $A\beta$ and $Ca^{+2}$ chosen here are $400\si{nM},0\si{nM}$, since the patients data has been fitted at $12$ month visit (Fig. \ref{fig:5}). The time scale chosen here is the age of patients ranging from $57-87$ years based on the ages of patients given on ADNI data.}
        \label{fig:6}
\end{figure}

In this section, we will present the results obtained from the developed physics-based Bayesian model presented in Section \ref{meth}. Our model can reproduce typical $A\beta$ and $Ca$ growth dynamics with or without the influence of stochastic noise. In the present study, we are interested in the analysis of such dynamics based on stochastic models of $A\beta$ and $Ca$ (i.e., Eqs. \ref{eq3}-\ref{eq4}), since AD models are predominantly stochastic. Moreover, we have fitted the data for $\mathbf{A\beta}$ for AD patients from the ADNI database within $2$-year visits in the developed stochastic model. We used the Bayesian inference approach to fitting stochastic differential equations to data. In Bayesian inference, a prior distribution is placed on the parameters, and the posterior distribution of the parameters is estimated using Bayes' theorem, which takes into account both the likelihood of the data and the prior distribution of the parameters. Once the parameters have been estimated, the equations can be used to simulate the concentrations of $\mathbf{A\beta}$ and $\mathbf{Ca}$ over time, which can be compared to the experimental data. The latter is a mathematical model of $\mathbf{A\beta}$ and $\mathbf{Ca}$ that has been formulated within a Bayesian setting to understand the $\mathbf{Ca}$ dynamics mediated by $\mathbf{A\beta}$ in AD. Note that in Figs. \ref{fig:4}-\ref{fig:6}, we produced the results based on the concentrations of $\mathbf{A\beta}$ and $\mathbf{Ca}$.

\begin{figure}[htbp]
\centering
\includegraphics[scale=.5]{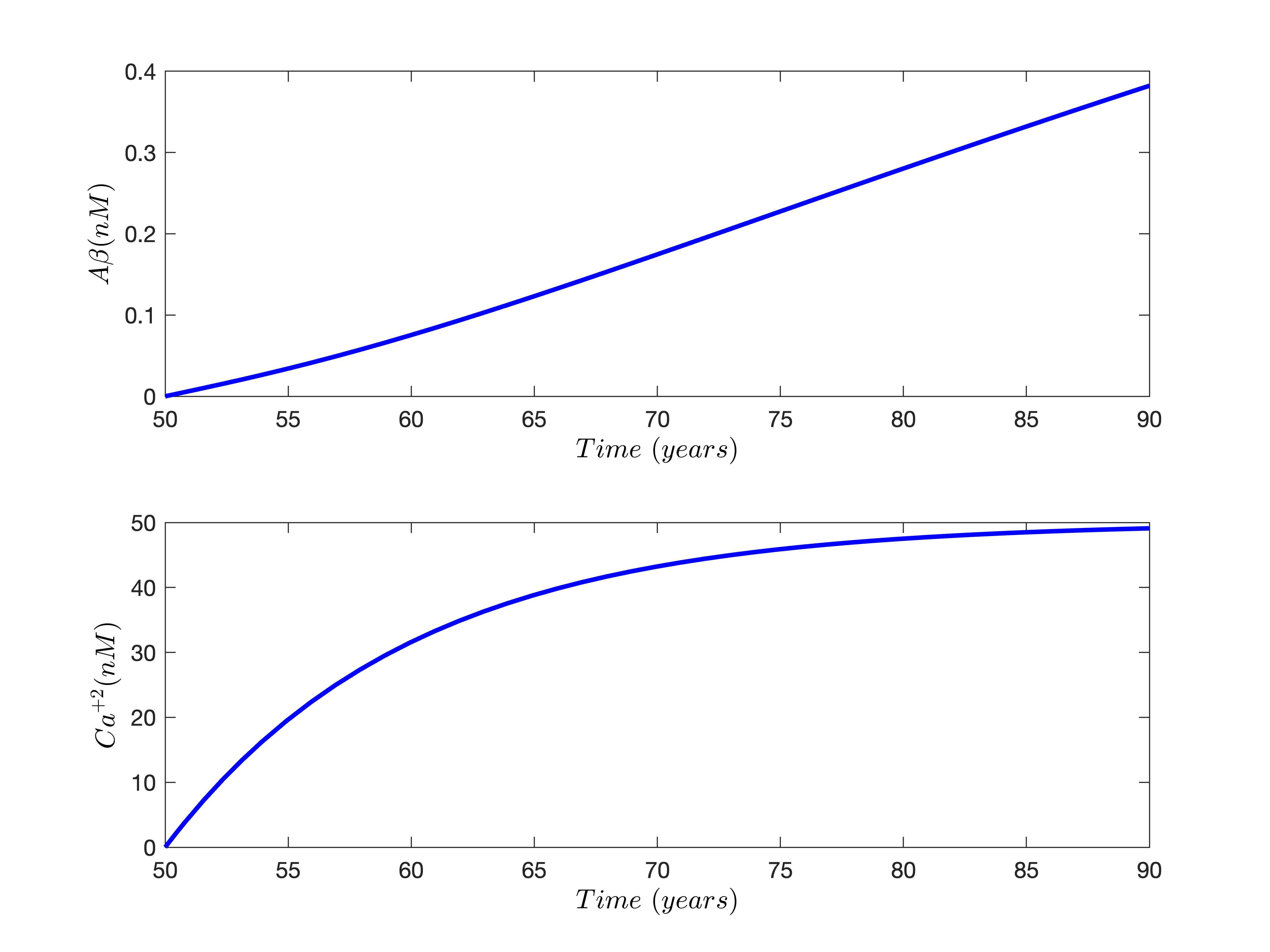}
\caption{(Color online) Simulation of the transformation from a healthy to a pathological state in the model for the onset of AD, specified by Eqs. (\ref{eq1})-(\ref{eq2}). The transition is caused by a shift in $Ca^{+2}$ homeostasis, which is represented in the model by increasing the rate of $Ca^{+2}$ entry ($V_2$) from $4$ to $5$. The initial concentration of both $A\beta$ and $Ca$ has been assumed to be zero, respectively. The parameter values are given in Table \ref{tab:1} and these values were chosen such that the $Ca^{+2}$ concentration, develops quicker than the amyloid $A\beta$ concentration given that the latter is defined by time scales on a number of years, whereas $Ca^{+2}$ is characterized by seconds to minutes (reproduced from \cite{de2013progression}).}
\label{fig:2}       
\end{figure}
To get an initial insight into the interplay between $A\beta$ and $Ca$, the simulation of the developed deterministic model (i.e., Eqs. (\ref{eq1})-(\ref{eq2}) onset of AD describing the positive loop between $A\beta$ and $Ca^{+2}$ is depicted in Fig. \ref{fig:2}. It can be seen that the model for the onset of AD simulates the shift from a healthy to a diseased state. Here, from an initial scenario corresponding to a stable state defined by low values of $A\beta$ and $Ca^{+2}$ concentrations represent the healthy situation. In contrast, the high values of $A\beta$ and $Ca^{+2}$ concentrations represent the pathological situation \cite{de2013progression}. According to the deterministic model (i.e., Eqs. (\ref{eq1})-(\ref{eq2}), the pathological state may only be attained by long-term changes in $Ca^{+2}$ homeostasis, including those that impact the $Ca^{+2}$ fluxes, as shown in Fig. \ref{fig:2}. Our findings suggest that when the basal level of $Ca^{+2}$ in the body increases over time, it can decrease the effectiveness of $\mathbf{Ca}$ signalling triggered by receptors \cite{hoffmann2003elevation}.

\begin{figure}[htbp]
     \centering
     \begin{subfigure}[b]{0.45\textwidth}
         \includegraphics[width=\textwidth]{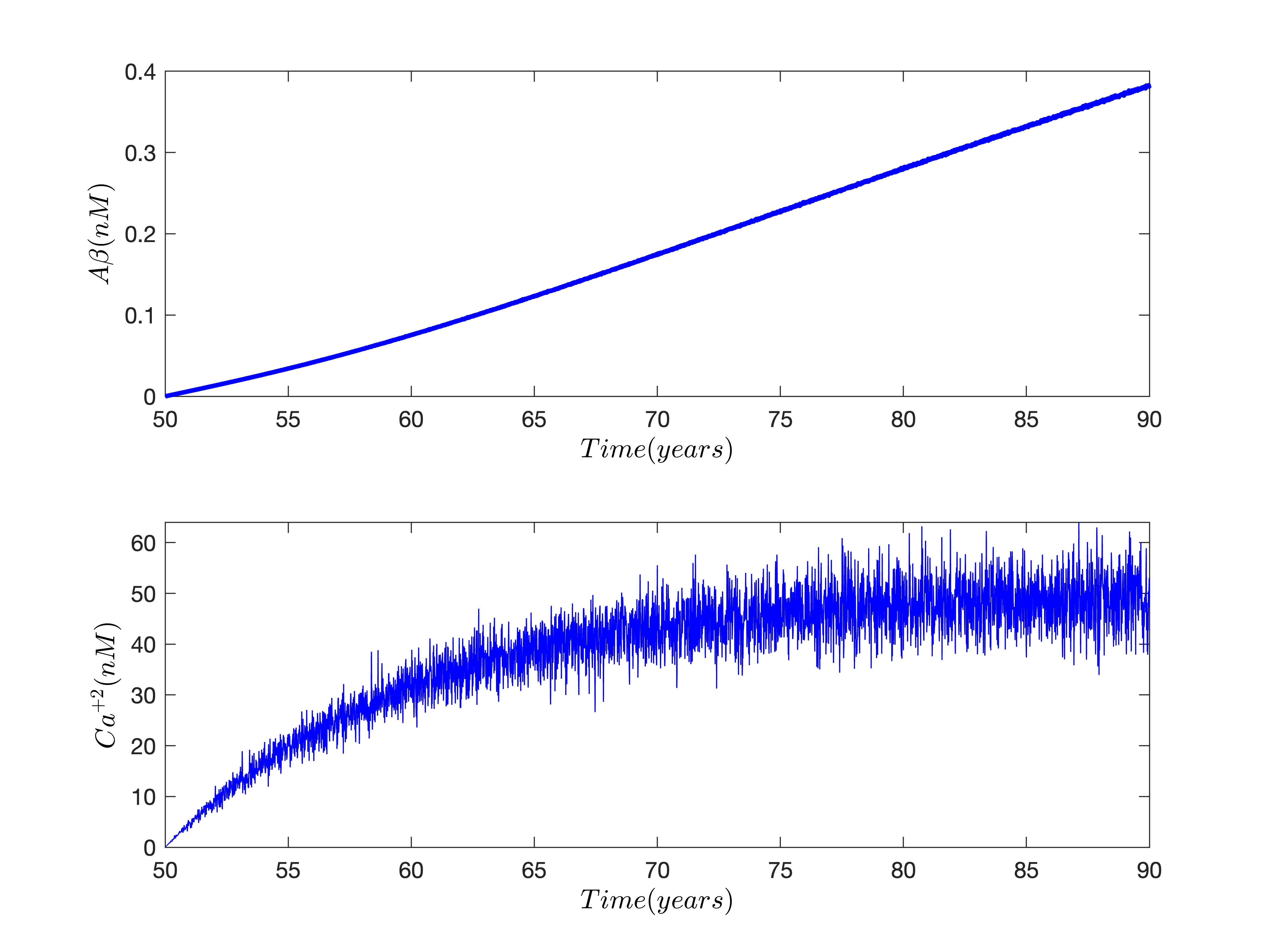}
         \caption{}
   \label{foura}
     \end{subfigure}
     \begin{subfigure}[b]{0.45\textwidth}
         \includegraphics[width=\textwidth]{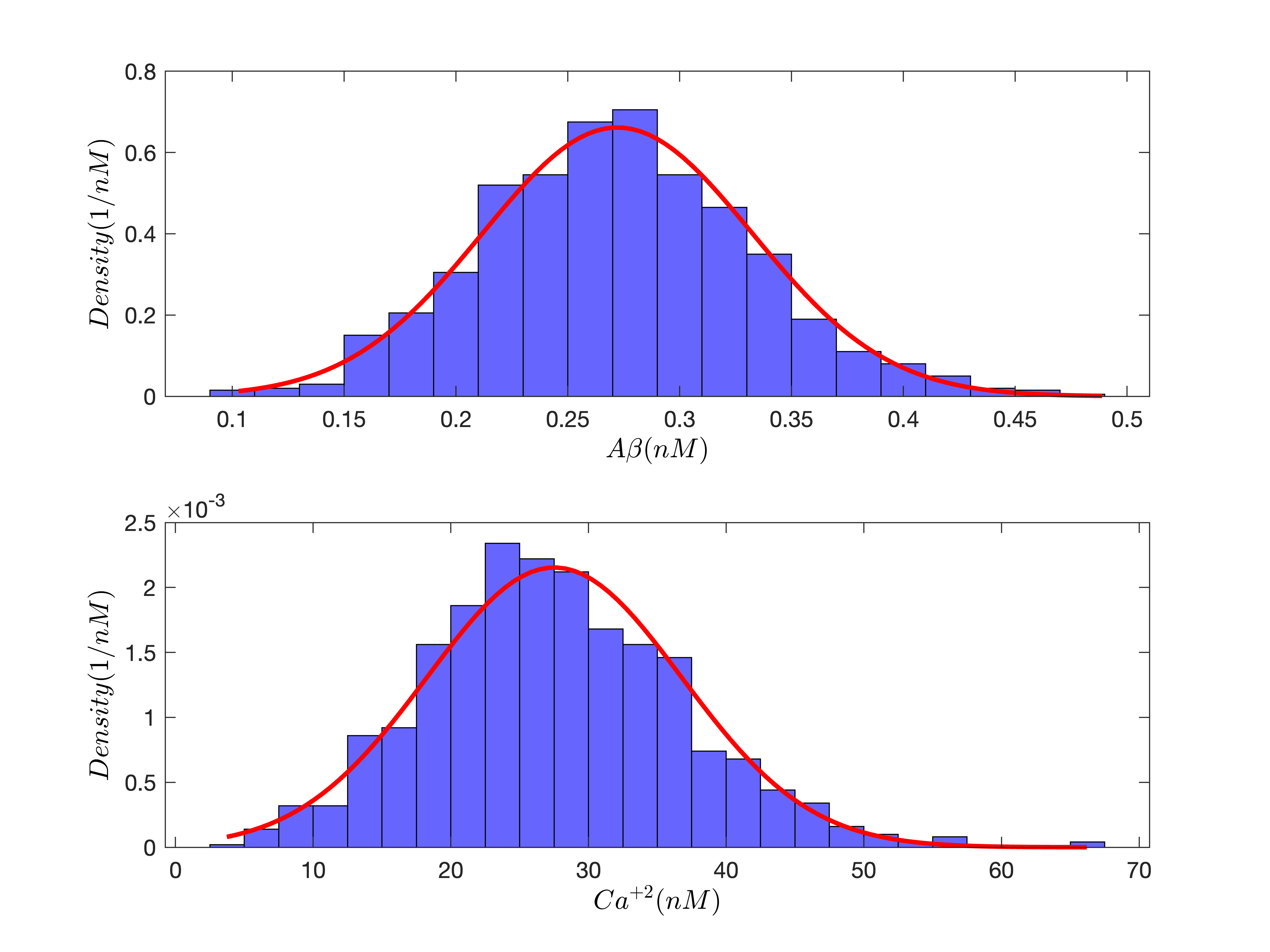}
         \caption{}
       \label{fourb}
     \end{subfigure}
        \caption{(Color online) (a) The concentration of $A\beta$ and $Ca^{+2}$ in the presence of the Wiener process (i.e., Eqs. (\ref{eq3})-(\ref{eq4}), where $\sigma_1 = 0.2;\epsilon = 0.01;\sigma_2 = 0.1$. (b) The blue histograms represent stationary density obtained by stochastic stimulations, the red lines stand for stationary density in Eqs. (\ref{eq3})-(\ref{eq4}), the y-axis represents the probability density function ($1/nM$). The relevant parameters are given in Table \ref{tab:1}.}
        \label{fig:3}
\end{figure}
Stochastic noises play a crucial role in the metabolism of $A\beta$ and $Ca^{+2}$ \cite{zhang2020mathematical}. From now on, we will report the results based on the stochastic model, i.e., Eqs. (\ref{eq3})-(\ref{eq4}), since the purpose of our study is to fit the ADNI data onto the developed stochastic model. Before fitting the ADNI data we will delineate the results using Eqs. (\ref{eq3})-(\ref{eq4}) based in stochastic simulations as presented in Fig.~\ref{fig:3} and Fig.~\ref{fig:4}. In Fig. \ref{fig:3} (a), we show that oscillations caused by molecular noise can cause a system to jump from the healthy state to the pathological steady state. Fig.~\ref{fig:3} (a) illustrates how such a variation might lead to the disease. It's interesting to note that while $Ca^{+2}$ evolves more quickly than $A\beta$, alterations in $Ca^{+2}$ alone are unlikely to cause disease. In other words, the shift to the pathological steady state cannot be reversed by a short-duration of $Ca^{+2}$ rise, which is a characteristic of the model that makes sense given that, without this feature, every action potential would lead to an increase in $A\beta$ as discussed in \cite{shaheen2021neuron,pal2022influence}. Yet, because of positive feedback and the rapid development of $A\beta$, any noise-induced rise in $A\beta$ will also generate an increase in $Ca^{+2}$, reinforcing the initial increase in $A\beta$. Moreover, in Fig.~\ref{fig:3} (b), we plot histograms representing stationary density obtained by stochastic simulations of the Eqs. (\ref{eq3})-(\ref{eq4}). We define the number of simulations and the length of each simulation as $1000$. The Euler-Maruyama method has been used to simulate the stochastic differential equations for $A\beta$ and $Ca^{+2}$ for a large number of iterations until the system reaches its stationary density \cite{mao2015truncated}. As depicted in Fig.~\ref{fig:3} (b), the system of stochastic differential Eqs. (\ref{eq3})-(\ref{eq4}) follows a normal distribution which shows that the simulated data fit well. We found that, when the noise strength $\sigma_1$ and $\sigma_2$ increases from zero, the number of extrema of stationary density obtained by stochastic simulations changes. By varying these two parameters, the analysis could investigate how the stationary state properties of the system change with different levels of noise strength. For example, the analysis could evaluate the number of extrema of the stationary density obtained by stochastic simulations, as well as other relevant properties such as the mean and variance of the system's state.

Note that in Figs.~(\ref{fig:2}-\ref{fig:3}), we produced the results based on the model Eqs. (\ref{eq1}-\ref{eq4}) without fitting the data. The next goal is to plot the predicted results by fitting the simulated data before adding the ADNI data into the Eqs. (\ref{eq3})-(\ref{eq4}) using Bayesian inference as described in Section \ref{abc}. Therefore, we set up the initial conditions accordingly. At first, the simulated stochastic trajectories were sampled for three different values of $A\beta_0$ for the stochastic model (i.e., Eqs. (\ref{eq3})-(\ref{eq4})) as presented in Fig.~\ref{fig:4}. These initial values of $A\beta$ are chosen based on the ADNI data (i.e., we choose the mean values of $A\beta$ concentrations at the bl, $12$ and $24$ month visits as the true initial value of $A\beta$, i.e., $A\beta_0$ for fitting ADNI data) and time is chosen in years that is actually the age of patients. So that the simulated trajectories corresponding to the time periods utilized in the experimental measurements of $A\beta$ completely align.
As predicted in Fig.~\ref{fig:4}, we fit the stochastic data using Eqs. (\ref{eq3})-(\ref{eq4}) for $A\beta$ concentration and analyzed its impact on $Ca^{+2}$ concentration. As evident from this figure, the temporal evolution of $A\beta$ concentration is increasing as $A\beta_0$ increases. In return, there is a large jump in the $Ca^{+2}$ concentration and it keeps increasing until the system goes to the disease state. This makes sense that increasing $A\beta$ concentration would increase $Ca^{+2}$ concentration, which leads to the progression of AD. Also, because of the positive feedback and the rapid growth of $A\beta_0$, any noise-induced rise in $A\beta$ will also produce an increase in $Ca^{+2}$, which will reinforce the increase in $A\beta$ \cite{shaheen2021neuron,zhang2020mathematical,calvo2020increased}. 

Importantly, it should be noted that the amount of $A\beta$ is a significant biomarker of AD. The National Institute on Aging (NIA), the Alzheimer's Association, and ADNI have recommended new criteria for diagnosing AD, and one of those suggestions uses $A\beta$ as a biomarker for an early diagnosis of AD \cite{bilgel2019predicting,thuraisingham2018pathogenesis,forsberg2008pet,koran2014genetic}. Medical studies have demonstrated that AD without symptoms can be found by counting the particles of $A\beta$ in the cerebrospinal fluid or brain \cite{mckhann2011diagnosis,albert2013diagnosis,sato2019lower}. As science and technology advance, several biomarker tests to gauge the level of $A\beta$, such as beta-amyloid PET imaging and cerebrospinal fluid testing, are being employed in some settings to support the diagnosis. Moreover, work is being done to create straightforward and affordable biomarker tests \cite{bilgel2019predicting,robin2018astroglial}. Therefore, based on this knowledge, the progression of disease associated with the amount of $A\beta$ used in this study helps predict the likelihood of developing AD because of the significant role that $A\beta$ plays in the diagnosis of AD and $Ca^{+2}$ dysregulation. Our aim is to fit the ADNI data for $A\beta$ concentration using the stochastic model (i.e., Eqs. (\ref{eq3})-(\ref{eq4})) and see the impact of $A\beta$ on $Ca^{+2}$ concentration for AD patients as per a $2$-year visit. In the context of solving coupled differential equations i.e., Eqs. (\ref{eq3})-(\ref{eq4}), ABC involves defining a model that generates the data, choosing a prior distribution for the model parameters, and simulating data from the model. The resulting posterior distribution can be used for inference about the model and the system being studied. Therefore, we have added the ADNI data and solved the inverse problem (i.e., Eq. (\ref{eq3})) using the ABC technique as discussed in Section \ref{abc}. Then the resulting data from Eq. (\ref{eq3}) has been added to Eq. (\ref{eq4}) to analyze the system for the interplay between $Ca^{+2}$ and $A\beta$ during the onset of AD. The two coupled equations i.e., Eqs. (\ref{eq3})-(\ref{eq4}) can be treated as an inverse problem by using Bayesian inference to estimate the values of the unknown parameters that give rise to the observed data. Specifically, we used the available $A\beta$ data on the ADNI database to estimate the disease progression trajectories for Eq. (\ref{eq3}), and then use these estimated trajectories as input to Eq. (\ref{eq4}). We can then simulate the biomarker measurements predicted by Eq. (\ref{eq4}) and compare them to the observed biomarker measurements. Then we estimated the posterior distribution of the model parameters given the observed data, since in an inverse problem, we are trying to infer the values of the unknown parameters. The corresponding $A\beta$ concentrations with relative contributions of the $Ca^{+2}$ concentrations are evaluated by Eqs. (\ref{eq3})-(\ref{eq4}) and shown in Fig.~\ref{fig:5}. The effects of different values of $A\beta_0$ on $A\beta$ concentrations at $bl$, $12$, and $24$ months visits are presented in Fig.~\ref{fig:5} (a). The time on the x-axis represents the age of patients per year, assuming that the participants are all $57$ years old at baseline as per the ADNI data. It can be seen that there is a rapid growth in the concentration of $A\beta$ for AD patients with three different frequency visits from bl to $24$ months. At an initial stage, AD patients have lower $A\beta$ concentrations, but they keep increasing with the passage of time. In return, there is rapid growth in the $Ca^{+2}$ concentrations, which corresponds to $A\beta$ growth during the AD as presented in Fig.~\ref{fig:5}(b). This shows that the inclusion of $A\beta$ has altered the steady-state populations of $Ca^{+2}$ due to the interactions between $A\beta$ and $Ca^{+2}$ signalling pathways as depicted in Fig.~\ref{fig:5}. $A\beta$ can disrupt calcium homeostasis by promoting calcium influx and inhibiting calcium efflux, leading to an increase in intracellular calcium levels. This increase in intracellular calcium can alter the steady-state populations of $Ca^{+2}$ by affecting the dynamics of calcium-dependent processes, such as calcium-dependent enzyme activation, gene expression, and synaptic plasticity \cite{alzheimer2017calcium}. Moreover, it is noteworthy that the $Ca^{+2}$ ions stimulate $A\beta$ production, which increases the $Ca^{+2}$ concentrations entering the cytoplasm of neuronal cells, resulting in a positive feedback loop. Interestingly, as the AD patient's age increases, there is a rapid growth in $A\beta$ and $Ca^{+2}$ concentrations since AD progresses. The results and the model predictions obtained in Fig.~\ref{fig:5} (a-b) align with previous theoretical and experimental studies, which reveal a feedback loop between $Ca^{+2}$ levels and $A\beta$ \cite{zhang2020mathematical,de2013progression,andrade2013cell,demuro2011single,kuchibhotla2008abeta,berridge2016inositol}. 

Next, to determine whether the ADNI data fit well in the stochastic models (i.e., Eqs. (\ref{eq3})-(\ref{eq4})) presented in Fig.~\ref{fig:5}, we plotted the histogram as shown in Fig.~\ref{fig:6} (a). As depicted in Fig.~\ref{fig:6} (a) after fitting the ADNI data, we see that both histograms for $A\beta$ (top) and $Ca^{+2}$ (bottom) follow the normal distribution. This demonstrates that the distribution of an ADNI data set follows a normal distribution, with the majority of data points clustering around the mean value and fewer data points in the tails of the distribution. The number of simulations and the length of each simulation (which is in years) are set to $1000$. The normal distribution proves that the ADNI data fits well into the Eqs. (\ref{eq3})-(\ref{eq4}). Since the data set follows a normal distribution, it allows us to make certain statistical inferences and predictions about the data. For example, we can use the mean and standard deviation of the distribution to calculate the probability of observing the experimental data and estimating the unknown parameters (i.e., $V_1$, $V_{\alpha}$, $K_{\alpha}$, $n$, $k_1$, $V_2$, $k_{\beta}$, and $m$) using the ABC technique as described in Section \ref{abc}. In the present study, using Bayesian inference, the likelihood function is used to quantify the probability of observing the ADNI data given a particular parameter value or set of parameter values, as shown in Table \ref{tab:1}. Specifically, the likelihood function is a function that takes in the observed data and a set of model parameters and returns the probability of observing the data given those parameter values \cite{held2020likelihood}. Thus, using this approach, the estimated parameters (i.e., $V_1$, $V_{\alpha}$, $K_{\alpha}$, $n$, $k_1$, $V_2$, $k_{\beta}$, and $m$) within the $95\%$ confidence interval are $0.00720,0.0435,125,1.98,0.02,4.989,0.0021,3.980,0.099$, respectively, by taking the true parameter values as given in Table \ref{tab:1}. Importantly, there is a difference in the time scales represented on the x-axis of Fig.~\ref{fig:3} (b) and Fig.~\ref{fig:6} (a). This distinction arises from the specific analysis conducted in each figure. In Fig.~\ref{fig:3} (b), we aimed to assess the goodness-of-fit between the simulated data, generated by stochastic simulations of $A\beta$ and $Ca^{+2}$ concentrations as illustrated in Fig.~\ref{fig:3} (a). On the other hand, in Fig.~\ref{fig:6} (a), we focused on fitting the patient data obtained at the 12-month visit (as shown in Fig.~\ref{fig:5} for $A\beta$ and $Ca^{+2}$ levels. The objective here was to see the stochastic simulation in the results of simulated data and whether ADNI data fit well or not taking into account the inherent stochasticity of the Wiener process used in the simulations and the time scales of the ADNI data. Therefore, due to the different analyses performed in Fig.~\ref{fig:6} (a) and Fig.~\ref{fig:3} (b), the histograms presented in Fig.~\ref{fig:3} (b) and Figure Fig.~\ref{fig:6} (a) have varying time scales.

Finally, the relation between corresponding $A\beta$ concentrations and the relative contributions of the $Ca^{+2}$ concentrations is shown in Fig.~\ref{fig:6} (b) after fitting the ADNI data. It can be seen that both concentrations are directly proportional to each other. This reveals, that there is positive feedback between $Ca^{+2}$ and $A\beta$. Specifically, as the concentration of $Ca^{+2}$ increases, it leads to an increase in the concentration of $A\beta$, which in turn leads to an increase in the concentration of $Ca^{+2}$, and so on. This positive feedback loop could be important in understanding the pathological mechanisms underlying AD, which is characterized by the accumulation of $A\beta$ in the brain. This is consistent with an experiment in which $Ca^{+2}$ levels are high during AD and there is positive feedback between  $Ca^{+2}$ levels and $A\beta$ concentrations \cite{zhang2020mathematical,de2013progression,andrade2013cell,demuro2011single,kuchibhotla2008abeta,berridge2016inositol,o2023calmodulin}. As a result of the bistable tendency \cite{de2013progression}, the final size of $A\beta$ is determined by the initial levels of $A\beta$ and $Ca^{+2}$. For example, if the initial concentrations of $A\beta$ and $Ca^{+2}$ are both set to low values, the system may not reach the threshold concentration needed to trigger the positive feedback loop, and the final size of $A\beta$ would be small. On the other hand, if the initial concentrations of $A\beta$ and $Ca^{+2}$ are both set to high values, the system may quickly reach the threshold concentration, and the positive feedback loop would be initiated, leading to a large final size of $A\beta$. By systematically varying the initial concentrations of $A\beta$ and $Ca^{+2}$ and monitoring the resulting final sizes of $A\beta$, one can observe the dependence of the final size on the initial levels of these concentrations and infer the existence of a bistable tendency\cite{itkin2011calcium,latulippe2018mathematical,zhang2020mathematical}. Another strategy in this direction is to assess the dynamics of $Ca^{+2}$ inside neurons as well as any existing alterations to its control during the development of the disease. It has been demonstrated experimentally that there may be a feed-forward loop between $A\beta$ and $Ca^{+2}$ regulation \cite{vosoughi2020mathematical}. Mathematically, this has been proven to be a bistable switch, in that when low levels of $A\beta$ and $Ca^{+2}$ (i.e. healthy state) begin to rise due to any form of disruption that leads to increased $A\beta$ or intracellular $Ca^{+2}$, resulting in certain pathologic effects \cite{de2013progression}. This suggests that if the existing levels of $A\beta$ and $Ca^{+2}$ are adjusted to their healthy states, the advancement of AD can be prevented. Because of the positive link between $Ca^{+2}$ and $A\beta$, this might be achieved by reducing $Ca^{+2}$ uptakes. The second technique to delay the development of the disease is to raise the intensity of stochastic noise for $Ca^{+2}$ and reinforce the strength of stochastic noise for $Ca^{+2}$ since these noises tend to lessen the severity of the disease when the stationary distribution is unimodal. They are made feasible by controlling the microenvironment for $A\beta$ and $Ca^{+2}$. To the best of our knowledge, these are novel results based on the association of $A\beta$ concentration obtained from ADNI data. The interactions between $A\beta$ and $Ca^{+2}$ add a new degree of complexity to important processes related to the beginning and progression of AD and may help to explain why viable therapeutic therapies for the disease have yet to be developed. 

The novelty of our present study is that the proposed model utilizes clinical data to measure the relationship between $A\beta$ and $Ca^{+2}$, specifically ADNI data gathered during per $2$-year visits with AD patients. The initial conditions for $A\beta$ and $Ca^{+2}$ are chosen based on our sole assumptions considering the specific data frame since we are only interested in the interplay between $A\beta$ and $Ca^{+2}$, once the disease started in the individuals. The data was analyzed using the ABC approach and fit our proposed model. The Markov chain Monte Carlo algorithm is used for the model that is fully coupled and for parameterizations. Our research has shown that the presence of $A\beta$ can create a zone with a steady state that is bistable. This means that the system can exist in two stable states, depending on the initial conditions. However, we used the ADNI data for AD patients only, so we discussed only the disease state and progression of AD. It is believed that the growth of cytosolic $Ca^{+2}$ due to $A\beta$ can lead to the development of AD. Our focus is on the quick development of abnormal $Ca^{+2}$ signals. The concentration of $A\beta$ changes over a much longer period than $Ca^{+2}$ due to accumulation over months, years or even decades. In this case, our research has shown that the presence of $A\beta$ in the model can create a zone with a bistable steady-state population of $Ca^{+2}$ due to the disruption of calcium homeostasis. However, the question of whether the steady-state population of $Ca^{+2}$ in AD patients is bistable requires further investigation and validation using experimental data. Multifidelity modelling is an approach used to efficiently predict the behaviour of a complex system by using multiple computational models of varying levels of fidelity or accuracy \cite{aydin2019general}. In the present study, we used the MCMC approach to combine low and high-fidelity calculations. In this approach, the low-fidelity calculation is used to generate a proposal distribution, which is then refined by the high-fidelity model. Next, the MCMC algorithm iteratively samples from the proposal distribution and accepts or rejects the samples based on their likelihood. This allows the low-fidelity calculations to explore the parameter space efficiently while the high-fidelity calculations are used to refine the results and increase their accuracy. The use of clinical datasets such as ADNI, in conjunction with the computational modelling described herein, facilitates the implementation of multi-fidelity association studies, which are novel and promising tools for evaluating the potential benefits and side effects of therapeutic agents that target known AD pathways. 
The study has the potential to generate novel ideas and hypotheses for future research, particularly in medication discovery and safety initiatives, which can be validated through additional cohort studies and clinical trials. Moreover, the findings may stimulate further research in the field by highlighting the significance of exploring $Ca^{+2}$ and their potential effects on $A\beta$.

The computational results have been obtained with an in-house developed MATLAB code to solve the coupled ordinary differential equations of the developed models (i.e., Eqs. (\ref{eq1}-\ref{eq4}) presented in the earlier Section \ref{meth}. Our data contains $1706$ patients, and we divide our job into many processors. We have selected the number of CPUs as the divisors of $1706$. For example, for $10,000$ iterations, for $1$ CPU, the computational time is $5012.1\si{s}$, and for $2$ CPU's it is $140.3\si{s}$, etc. To attain the required total time for each model, it often takes several million-time steps. If we solve the problem using conventional serial programming, this requires a significant amount of computing time. Through the use of open MPI and the C programming language, we can reduce computing time. We divide the sequential tasks involving the time step among available processors for each time iteration and perform them in parallel. All figures have been plotted and shown in Matlab after the data have undergone post-processing. To reduce the time required to acquire results for the parallel computation, we employed the SHARCNET supercomputer facilities (64 cores).
 
\section{Discussion}\label{dis}
In this paper, we extended the stochastic mathematical model of AD by introducing ADNI data for $A\beta$ and analyzed the interplay between  $A\beta$ and $Ca^{+2}$. We investigated the dynamical behaviours of stochastic processes with the model by incorporating slow-fast timescales between $A\beta$ and $Ca^{+2}$, which revealed the influence of random noises on the advancement of AD. The number of AD modelling tools available to date has been fairly limited, most likely due to the enormous complexity of the molecular systems underlying its pathogenesis. Many mathematical models have previously been constructed to explore particular and well-defined features of the disease \cite{pal2022influence,pal2022coupled,fornari2020spatially,fornari2019prion}. To the best of our knowledge, none of the computational models have been proposed yet to explore the synaptic interplay between $A\beta$ and $Ca^{+2}$ using clinical data. Here, we provide a simple stochastic model that qualitatively describes the interactions between intracellular $A\beta$ and $Ca^{+2}$ using ADNI data. It is based on two simple coupled stochastic  differential equations with the Wiener process or stochastic noises.

Our goal was to analyze any potential functional effects or interplay between $A\beta$ and $Ca^{+2}$ resulting from the positive loop that exists between the two chemicals in AD patients' brains, despite the fact that the model is obviously oversimplified. Importantly, it is possible for $A\beta$ to bind to NMDA receptors and produce $Ca^{+2}$ dyshomeostasis, which results in oxidative stress, the formation of free radicals, and the death of neurons \cite{john2021synaptic}. Furthermore,  $A\beta$ can activate mGluR5 receptors, which elevates postsynaptic $Ca^{+2}$ levels in the cell \cite{shaheen2021neuron,sheng2012synapses}. Then, APP processing is accelerated by NMDA receptors and mGluR, which create a positive feedback loop that boosts $Ca^{+2}$ influx and free radical generation \cite{rajmohan2017amyloid}. We have shown that it explains well-known aspects of the disease, such as its inability to be reversed, the threshold-like transition to severe pathology following the relatively slow accumulation of symptoms, the so-called ``prion-like" autocatalytic behaviour, and the naturally random nature of the disease's emergence that is typical of AD in sporadic cases. 

Nonetheless, there are several more general characteristics of bistable behaviour that may be mentioned. Here, the bistable behaviours mean that the final size of $A\beta$ is dependent on the initial levels of $A\beta$ and $Ca^{+2}$ \cite{zhang2020mathematical}. First, since each neuron is either in one steady state or the other, with those two states being distinguished by very different values for concentration levels of $A\beta$ and $Ca^{+2}$and enzymatic activities, average measurements of $A\beta$ or $Ca^{+2}$-related quantities are expected to have little significance in terms of experimental observations on disease characteristics. The model contends that comprehensive identification of the condition of the neurons, or at least of the damaged ones, is required for experimental quantification. Therefore, in the present study, we considered only AD patients, and then in damaged neurons, we analyzed the interplay between $A\beta$ and $Ca^{+2}$. Our previous study \cite{shaheen2021neuron} shows that $A\beta$ enhances the dysregulation of $Ca^{+2}$ in AD, so the present studies validate our hypotheses. Furthermore, in terms of the progression of Alzheimer's disease, it is intriguing to note that dysregulations in both $A\beta$ and $Ca^{+2}$ constitute the causes and effects of the disease in this scenario originating from the positive loop. In this way, it unifies two theories for the onset of AD that are frequently presented as opposed in the literature: the ``amyloid hypothesis", in which $A\beta$ is presented as the causative factor, and the ``$Ca^{+2}$ hypothesis", in which the up-regulation of $Ca^{+2}$ signalling is assumed to play the primary role. Both molecules are intimately connected to one another and equally responsible for the development of AD during the clinical trial, according to the current study.

Our model is limited by the simplification of a single pathway of $A\beta$ changes from many stages (e.g., NL to MCI to AD using ADNI data) to AD. It is well recognized that AD progresses in many ways throughout the years. If large patient samples are researched over an extended period of time, clinical trials are difficult to conduct and expensive. Moreover, modelling and simulating AD dynamics can be done for a small cost, and they are invaluable resources for improving clinical trial designs and raising the probability of accurate treatment efficacy assessments. The results obtained here are consistent with earlier theoretical and experimental investigations \cite{shaheen2021neuron,pal2022coupled,zhang2020mathematical, de2013progression,hao2016mathematical,andrade2013cell,demuro2011single,kuchibhotla2008abeta,berridge2016inositol,o2023calmodulin}. Our goal in this work was to describe the progression of AD, which encompasses not just AD pathology but also biochemical and cognitive alterations brought on by $A\beta$ and $Ca^{+2}$. The modelling strategy  developed here can calculate individual $A\beta$ and $Ca^{+2}$ growth trajectories and markers of latent disease progression at the population level using ADNI data. Individuals are identified along simulated trajectories by utilizing the proposed framework offered by Bayesian inference. This study highlights the rising role of $Ca^{+2}$ ions in the development of AD and focuses on the key components of the interplay between $A\beta$ and $Ca^{+2}$ homeostasis.

\section{Conclusions}\label{con}
Millions of individuals suffer from the progressive neurological disorder known as Alzheimer's disease. AD patients suffer from gradual, permanent cognitive deterioration. The biggest risk factor for AD is age. The development of plaques in the brain, caused by the gradual deposition of cerebral amyloid-$\beta$ ($A\beta$) peptides in the extracellular space, and of intracellular neurofibrillary tangles, made of misfolded proteins that typically stabilize microtubules with neuronal axons, are pathological symptoms of AD. Moreover, there is growing evidence that long-term disturbances of intracellular $Ca^{+2}$ homeostasis may be a key factor in AD. Particularly, it appears that $A\beta$'s cause an increase in intracellular $Ca^{+2}$ since multiple studies have discovered $Ca^{+2}$-dysregulations resulting in higher $Ca^{+2}$ entry in the cytoplasm in AD mouse models. Despite extensive studies, the pathophysiology of AD is still poorly understood, and the associated underlying molecular alterations have not yet been fully discovered. 

In the present study, we developed a simple, yet effective, stochastic model formalizing
a positive feedback loop between $A\beta$ and $Ca^{+2}$. The novelty of the proposed model is that it incorporates clinical data, such as ADNI data for AD patients per $2$-year visit, for quantifying the interplay between $A\beta$ and $Ca^{+2}$. The data were fitted to the given model using the ABC technique. The goal was to analyze the specific roles of $A\beta$ and $Ca^{+2}$  on synaptic homeostasis and discuss therapeutic protocols to slow down the progression of AD. 
More specifically, we investigate the underlying mechanisms that lead to neuronal hyperactivity and the role of $A\beta$ growth on the $Ca^{+2}$ dynamics. We demonstrated that in the AD brain, increasing $A\beta$ concentrations could lead to an increase in $Ca^{+2}$ dysregulation, which is harmful and promotes neuronal death. Moreover, there exists a positive feedback loop between the growth of both compounds (i.e., $A\beta$ and $Ca^{+2}$). It is expected that the proposed model will assist in a more precise prediction of the synaptic mechanism during AD and pave the way for the experimental testing of different hypotheses. We provided numerical simulations that agree with the previous findings that a number of dysregulations within the brain can lead to a disease state \cite{shaheen2021neuron,pal2022coupled,hao2016mathematical,andrade2013cell,demuro2011single,kuchibhotla2008abeta,berridge2016inositol,de2013progression,zhang2020mathematical,o2023calmodulin}. Importantly, changing the balance between $A\beta$ and $Ca^{+2}$ concentrations or lowering both concentrations may be able to alleviate AD-related disorders and open up new research avenues for AD treatment. Our findings fill gaps in AD research by explaining how $A\beta$ plaques develop, what happens when $Ca^{+2}$ and $A\beta$ interact, and how they induce selective neuronal death in AD patients. Future research will be able to more precisely evaluate model predictions and forecast the progression of the diseaseby including more patients, for instance, MCI, EMCI, and NL, to clinical trial data.

\section*{Acknowledgements}
The authors are grateful to the NSERC and the CRC Program for their support. RM is also acknowledging the support of the BERC 2022–2025 program and the Spanish Ministry of Science, Innovation and Universities through the Agencia Estatal de Investigacion (AEI) BCAM Severo Ochoa excellence accreditation SEV-2017–0718. This research was enabled in part by support provided by SHARCNET \url{(www. sharcnet.ca)} and Digital Research Alliance of Canada \url{(www.alliancecan.ca)}.

Data collection and sharing for this project was funded by the Alzheimer's Disease Neuroimaging Initiative (ADNI) (National Institutes of Health Grant U01 AG024904) and DOD ADNI (Department of Defense award number W81XWH-12-2-0012). ADNI is funded by the National Institute on Aging, the National Institute of Biomedical Imaging and Bioengineering, and through generous contributions from the following: AbbVie, Alzheimer's Association; Alzheimer's Drug Discovery Foundation; Araclon Biotech; BioClinica, Inc.; Biogen; Bristol-Myers Squibb Company; CereSpir, Inc.; Cogstate; Eisai Inc.; Elan Pharmaceuticals, Inc.; Eli Lilly and Company; EuroImmun; F. Hoffmann-La Roche Ltd and its affiliated company Genentech, Inc.; Fujirebio; GE Healthcare; IXICO Ltd.; Janssen Alzheimer Immunotherapy Research $\&$ Development, LLC.; Johnson $\&$ Johnson Pharmaceutical Research $\&$ Development LLC.; Lumosity; Lundbeck; Merck $\&$ Co., Inc.; Meso Scale Diagnostics, LLC.; NeuroRx Research; Neurotrack Technologies; Novartis Pharmaceuticals Corporation; Pfizer Inc.; Piramal Imaging; Servier; Takeda Pharmaceutical Company; and Transition Therapeutics. The Canadian Institutes of Health Research is providing funds to support ADNI clinical sites in Canada. Private sector contributions are facilitated by the Foundation for the National Institutes of Health \url{(www.fnih.org)}. The grantee organization is the Northern California Institute for Research and Education, and the study is coordinated by the Alzheimer's Therapeutic Research Institute at the University of Southern California. ADNI data are disseminated by the Laboratory for Neuro Imaging at the University of Southern California.

\bibliographystyle{elsarticle-num}
\bibliography{references}

\end{document}